\documentclass[10pt,twocolumn]{revtex4}

\usepackage{graphicx}
\usepackage{amsmath}

\newcommand{\bc}{{\bf c}}
\newcommand{\bm}{{\bf m}}
\newcommand{\bx}{{\bf x}}
\newcommand{\bv}{{\bf v}}
\newcommand{\bg}{{\bf g}}

\begin{document}

\title{Quantum Monte Carlo for minimum energy structures}

\author {Lucas K. Wagner and Jeffrey C. Grossman } 
\affiliation { 
Department of Materials Science and Engineering, \\ 
Massachusetts Institute of Technology, Cambridge, MA 02139 \\
E-mail:  lkwagner@mit.edu, jcg@mit.edu
}

\date{\today}

\begin{abstract}
We present an efficient method to find minimum energy structures using energy estimates from accurate quantum 
Monte Carlo calculations.  This method involves a stochastic process
formed from the stochastic energy estimates from Monte Carlo that can be averaged to find precise structural minima while using inexpensive calculations with moderate statistical uncertainty.  We demonstrate 
the applicability of the algorithm by minimizing the energy of the H$_2$O-OH$^-$ complex and showing that 
the structural minima from quantum Monte Carlo calculations affect the qualitative behavior of the potential energy 
surface substantially.
\end{abstract}


\maketitle 

%\section{Introduction}

First principles electronic structure methods have been used to 
describe and explain a wide range of properties for different condensed matter systems.  
A critical step is the  accurate determination of the ground 
state atomic structure, since many important properties of a material can change dramatically depending 
on the structure.
 Because of the balance between accuracy and computational cost,
  density functional theory (DFT) has become a commonly used method to find equilibrium geometries of 
both molecules and extended systems.  The primary reason for this is the availablility of 
forces with little extra computational cost over the energy calculation.  Using a typical quasi-Newton minimization algorithm, 
the local minimum of a potential energy surface can be found in ${\cal O}(N_{DOF})$, where  $N_{DOF}$ is the number of degrees of freedom to be optimized.  This favorable scaling has 
made it possible to find minima for many systems of interest.  However, in many situations, including transition
metals, excited states, and weak binding, current density functional theories may not be accurate enough even for 
structures, and more accurate
post-Hartree-Fock methods that scale as ${\cal O}(N_e^{5-7})$, where $N_e$ is the number of electrons, can often be too computationally expensive.

Quantum Monte Carlo (QMC), a stochastic approach to solving the many-body Schr\"odinger
equation, offers favorable scaling to obtain the total energy, ${\cal O}(N_e^{2-3})$, and has been shown to provide 
near chemical accuracy in many different systems\cite{jindra_feo, jeff_benchmark}.  However, there are two major challenges in using QMC methods
to obtain high precision minimum energy structures.  The first is that the techniques so far proposed to calculate  forces in diffusion Monte Carlo all have large variances and
error terms that depend on the quality of the trial wave function, which is often poor in systems where
DFT fails and one would like to apply QMC methods.
In fact, despite much work in recent years\cite{chiesa_force,assaraf_force, mella_force, badinski_force1, badinski_force2,filippi_correlated}, 
QMC forces using the highly accurate diffusion Monte Carlo method  have not been applied to more than a 
few degrees of freedom, although the simpler variational Monte Carlo technique has been applied to more\cite{attaccalite_md}.    The second challenge is the stochastic nature
of Monte Carlo algorithms, which provides uncertainty in any estimator that only decreases as
the square root of the computer time.  Reducing the uncertainty enough to resolve the minimum structure accurately using forces or total energies is often prohibitively expensive computationally.
Methods such as the stochastic gradient approximation\cite{monro, harju_sga} that are able to operate in the presence of noise suffer from this large uncertainty in the forces.  As a result, there are no geometry optimizations of more than three\cite{lester_fit} degrees of freedom, to our knowledge.

In this article, we describe an algorithm that uses the already-accurate total energies from QMC to obtain minimum energy 
geometries with well-defined stochastic uncertainties with multiple degrees of freedom.  The algorithm consists of two major parts.  One is a sequence of minimizations along one dimensional lines.  The use of 1D minimizations allows us to use efficient fits to determine the minimum precisely.  The second part is a quadratic fit of the many-dimensional energy surface to determine the new search directions.  Both of these parts are completely aware of the stochastic uncertainty present in the evaluations of the energy, an important feature obtained by the use of Bayesian inference.  We apply this approach to the hydrogen-transfer model of H$_2$O-OH$^-$ and show that our method can help clarify challenging problems that require accurate calculation of the electronic ground state.

%\section{Methods}

DFT and Hartree-Fock calculations were performed using the GAMESS\cite{gamess} package.  We used soft pseudopotentials\cite{Dolg_psp_qmc} 
to remove the core electrons and a TZP-quality contracted gaussian basis to represent the one-particle orbitals.
All-electron calculations were performed using  the aug-cc-pVQZ\cite{dunning_basis} basis to check the basis 
set and pseudopotential errors.
All QMC calculations were performed using the QWalk\cite{qwalk} program.  For energies, we used fixed node diffusion Monte Carlo (DMC) with a
time step of 0.02-0.05 Hartrees$^{-1}$, converged for the properties of interest.  The trial function was
a Slater-Jastrow function with hybrid PBE0\cite{pbe0} orbitals, one and two body terms in the Jastrow, and further checks of the localization error with a three-body Jastrow factor.  Further details can be found in 
e.g. Refs \cite{Foulkes_review, qwalk}.

\begin{figure}
\includegraphics[width=\columnwidth]{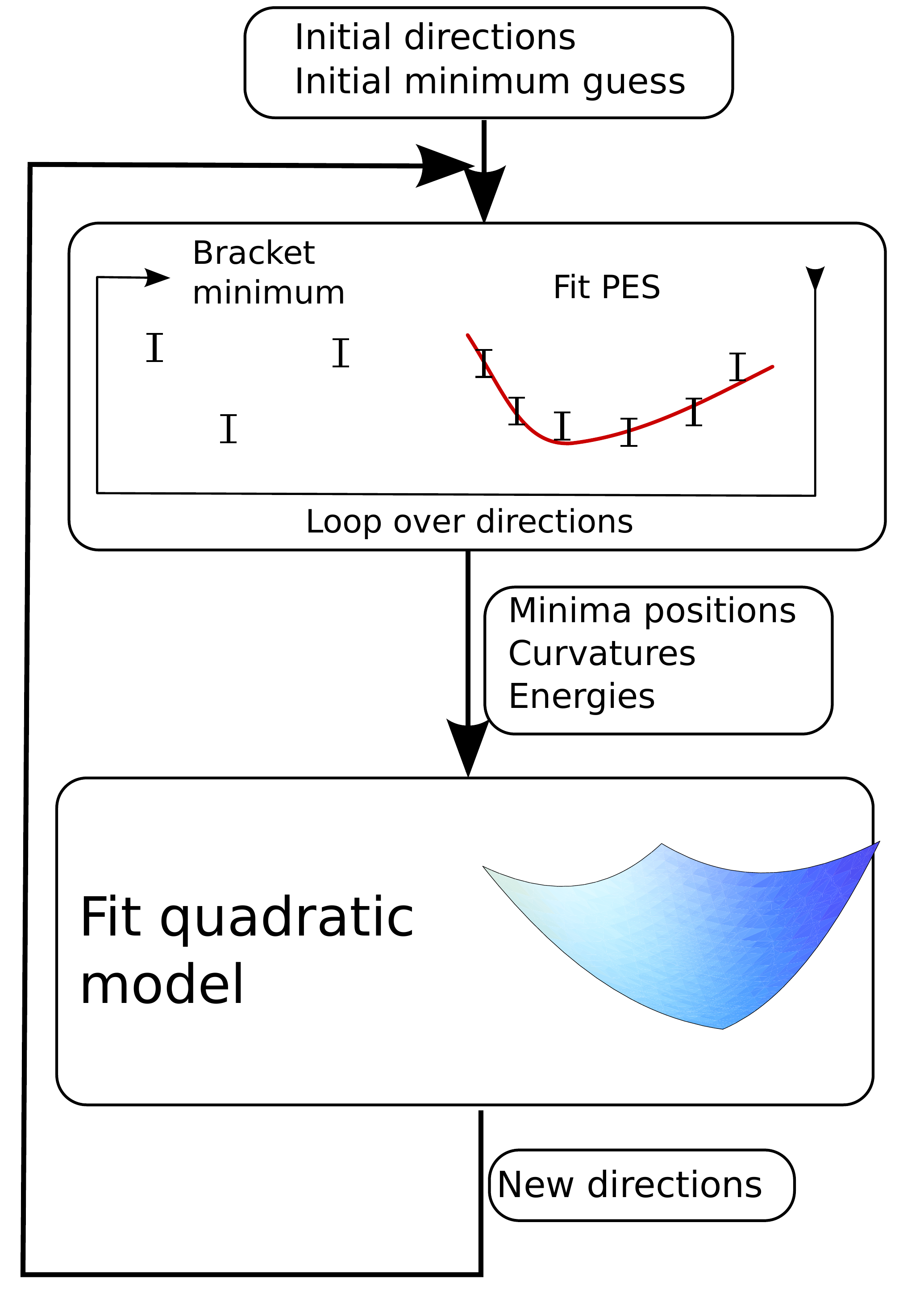}
\caption{(color online) An outline of the algorithm.  See text for details.}
\label{fig:algorithm}
\end{figure}

Our minimization algorithm is similar in spirit to many successful minimization algorithms in that it is based on minimizing along directions and updating an estimate of the Hessian matrix to find the diagonal directions.  However, in our approach the Hessian matrix is inferred using the Bayesian interpretation of probability, which has the effect of making the algorithm very robust to stochastic uncertainty.  Here we present the case when only the value can be calculated, but gradients of the objective function can be included easily if they are available, increasing the efficiency.  By using inference, we are able to make very efficient use of the available data to find the Hessian matrix without having to reduce the stochastic uncertainties to small values that cost large amounts of computational time to achieve.

The kernel of the algorithm is a sequence of line minimizations.  We show that a sequence of uncertain line minimizations
will obtain the true minimum on average.  Such a sequence can then be viewed as a generator of random variables whose 
expectation value is the true minimum.  Suppose
on the first step, we start with the approximate minimum $\bx^{(0)}$.  We then define $\delta \bx^{(0)}=\bx^{(0)}-\bm$, where
$\bm$ is the vector of the unknown true minima.  We wish to design our process such that 
the expectation value of  $\delta \bx^{(n)}$ equals zero as $n \to \infty$. 
Within the quadratic region around the minimum,  the potential energy surface is 
given by 
\begin{equation}
E(\bx)=\delta \bx^T H \delta \bx+E_0,
\end{equation}
where $H$ is the symmetric Hessian matrix and $E_0$ is the minimum total energy.
On minimizing along each direction $i$, there are two components of the distance from the true minimum.  The first 
is deterministic and comes about from non-zero $\delta x_j$ for $j\neq i$.  The second is the stochastic error from 
the uncertainty in the line minimization, which we can estimate using the Bayesian techniques above.
We find that 
\begin{equation}
\delta x_i^{(1)}=\chi_i^{(1)}+\sum_{j \neq i}\frac{H_{ij}}{H_{ii}}\delta x_j^{(0)},
\label{eqn:minimization_error}
\end{equation}
where $\chi_i^{(1)}$ is a random number.  For the $n$th iteration,
\begin{equation}
\delta x_i^{(n)}= \chi_i^{(n)}-\sum_{j \neq i} \frac{H_{ij}}{H_{ii}} \chi_j^{(n-1)} %\notag 
+\sum_{j \neq i} \sum_{k \neq j} \frac{H_{ij}H_{jk}}{H_{ii}H_{jj}} \chi_k^{(n-2)}+\ldots
\label{eqn:avg}
\end{equation}
The minimum along each line is found using line fitting, as described in the EPAPS document. This allows for a very efficient determination of the minimum.

One can see from Eqn~\ref{eqn:avg} that the smaller the off-diagonal matrix elements of the Hessian are, the less 
interference directions have on each other--for a diagonal Hessian, only one minimization in each direction is necessary 
once we are in the quadratic regime.  We can use the information from the line minimizations to estimate the 
Hessian as the minimization proceeds.  We parameterize the quadratic region with a set of parameters $\bc_Q$, including the 
elements of the Hessian matrix, the minima, and the minimum energy.  After performing the line fits, we have distributions of line fit parameters, each given by $\bc_\ell$ for line $\ell$. Using Bayes' theorem, the likelihood function of the quadratic parameters $L$ is given by 
\begin{equation}
 L(\bc_Q |D ) \propto p(D|\bc_Q) = \prod_\ell \int p(D_\ell | L_\ell) p(L_\ell | \bc_Q) d \bc_\ell ,
\end{equation}
 where D is the set of function evaluations and stochastic uncertainties given by e.g. QMC and each $D_\ell$ is the subset of function evaluations used to minimize along a given line $\ell$.  Since $p(L_\ell | \bc_Q)$ is not based on stochastic data, it is a delta function that forces $\bc_\ell$ to be consistent with $\bc_Q$ as follows.  Since in the quadratic region, $E(\bx)=\delta \bx^T H \delta \bx +E_0$, minimizing along a direction $\bv$ from a starting position $\bx_0$ gives the one-dimensional function of $t$, the position along the line:
\begin{equation}
E(t)=(\bx_0+t\bv-\bm)^T H (\bx_0+t\bv-\bm)+E_0.
\end{equation}
This gives the following constraints on each set of line parameters for the minimum, curvature, and minimum function value $c_\ell$: 
\begin{equation}
t_m=- \frac{ \bv^T H (\bx_0-\bm) + (\bx_0-\bm)^T H \bv }{ 2 \bv^TH\bv }
\end{equation}
\begin{equation}
\left.\frac{d^2E}{dt^2} \right|_{t_m} = 2 \bv^T H \bv
\end{equation}
\begin{equation}
E(t_m)= (\bx_0+t_m\bv -\bm)^T H (\bx_0 + t_m \bv - \bm)
\end{equation}

The net result of this transformation is a set of parameters that includes the Hessian matrix, the location of the minimum, the objective function at the minimum, and 
the parameters used for the line fits above the three constraints implied by the $\bc_Q$'s.  We examine the properties of this probability distribution function in the EPAPS material.  It turns out that the maximum likelihood estimator is typically accurate enough to determine the Hessian, which is then diagonalized to obtain new search directions.

\begin{figure}
\includegraphics[width=\columnwidth]{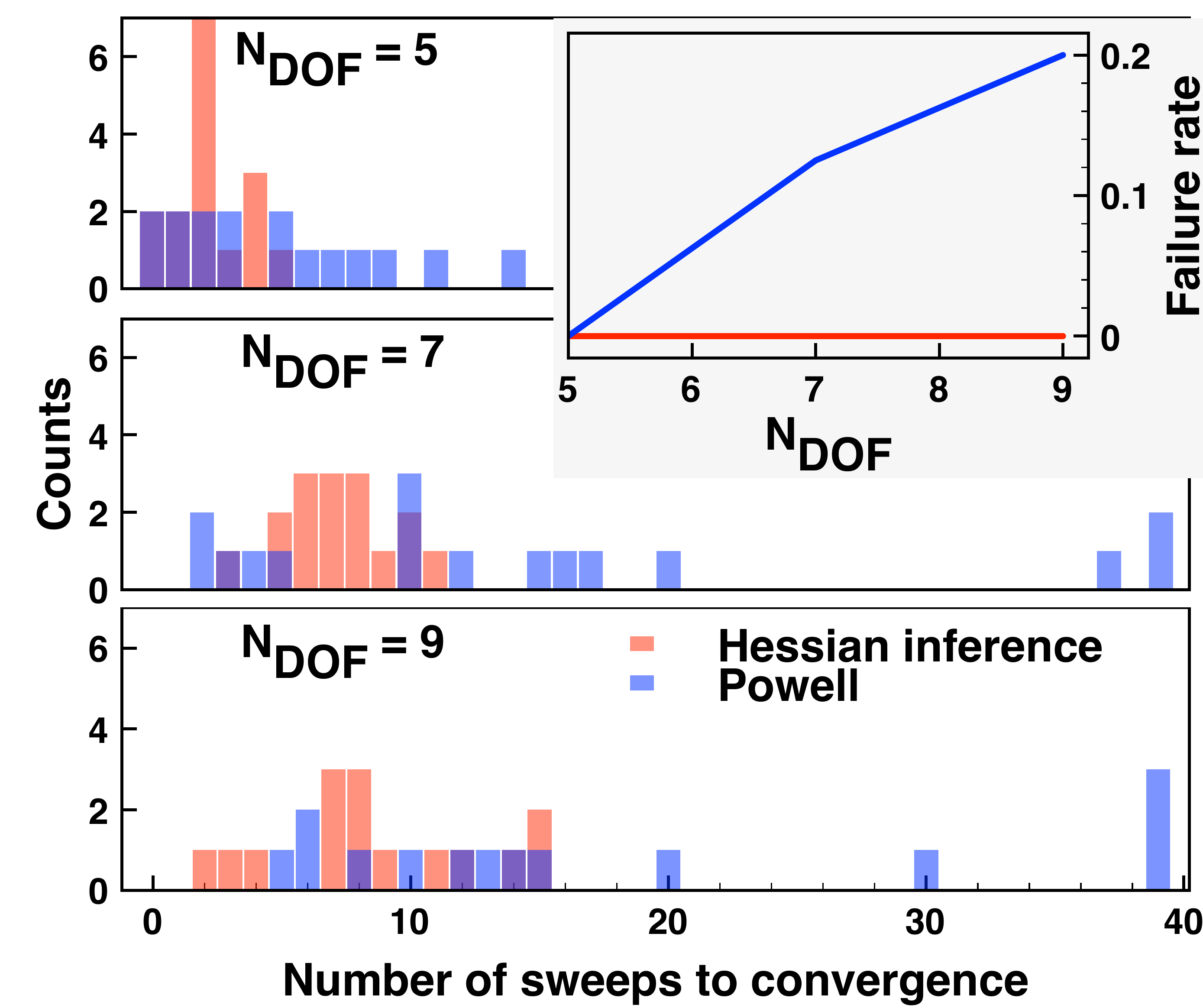}
\caption{(color online) Estimates of reliability for the Hessian inference method versus Powell line minimization.   There are several runs for each number of degrees of freedom, each with a different potential energy surface and relative starting position.  The potential surface was a randomly generated positive definite quadratic surface with a random minimum and a small cubic perturbation.  The condition number of the matrices was typically 50 or less.  We added a Gaussian random variable to the values of the potential energy surface to simulate uncertain evaluation. Convergence was judged when the RMS distance to the true minimum converged to the stochastic floor of the minimization.}
\label{fig:performance}
\end{figure}

We outline a single iteration of the algorithm in Fig~\ref{fig:algorithm}.  This step operates in two principle regimes.  The first is far away from the minimum.  In this regime, the Hessian inference
method behaves similarly to a deterministic direction set method, with the directions being determined by the Hessian inference.  
The second regime is when the calculation has mostly converged and the deterministic error in Eqn~\ref{eqn:minimization_error} is small compared to the stochastic error.  In this regime, deterministic direction choices are particularly useless, but the inferred Hessian method is able to automatically account for the stochastic error.  Once in this ``stochastic regime,'' we can use Eqn~\ref{eqn:avg} to justify averaging the $\bx^{(n)}$'s to obtain a more precise estimate of the minimum.
For stochastic functions, the performance of the Hessian inference is much higher than traditional methods such as Powell's method\cite{powell}, which we compare in Fig~\ref{fig:performance}.  Since Powell's method uses the concept of conjugate directions to find the search directions, the uncertain minimizations cause the algorithm to fail quickly.  Even when the directions are reset every sweep, they can become linearly dependent within a single sweep, causing a high failure rate for even moderate dimensionality.

%\begin{figure}
%\begin{tabular}{|c|c|}
%\hline 
%\includegraphics[width=4cm]{h3o2/noncentro.png}  & \includegraphics[width=4cm]{h3o2/centro.png} \\
%{\bf A } & {\bf B} \\
%\hline
%\end{tabular}
%\caption{(color online) Two candidate structures for the ground state of the H$_2$O-OH$^-$ complex.}
%\label{fig:h3o2}
%\end{figure}

\begin{figure}
\includegraphics[width=\columnwidth]{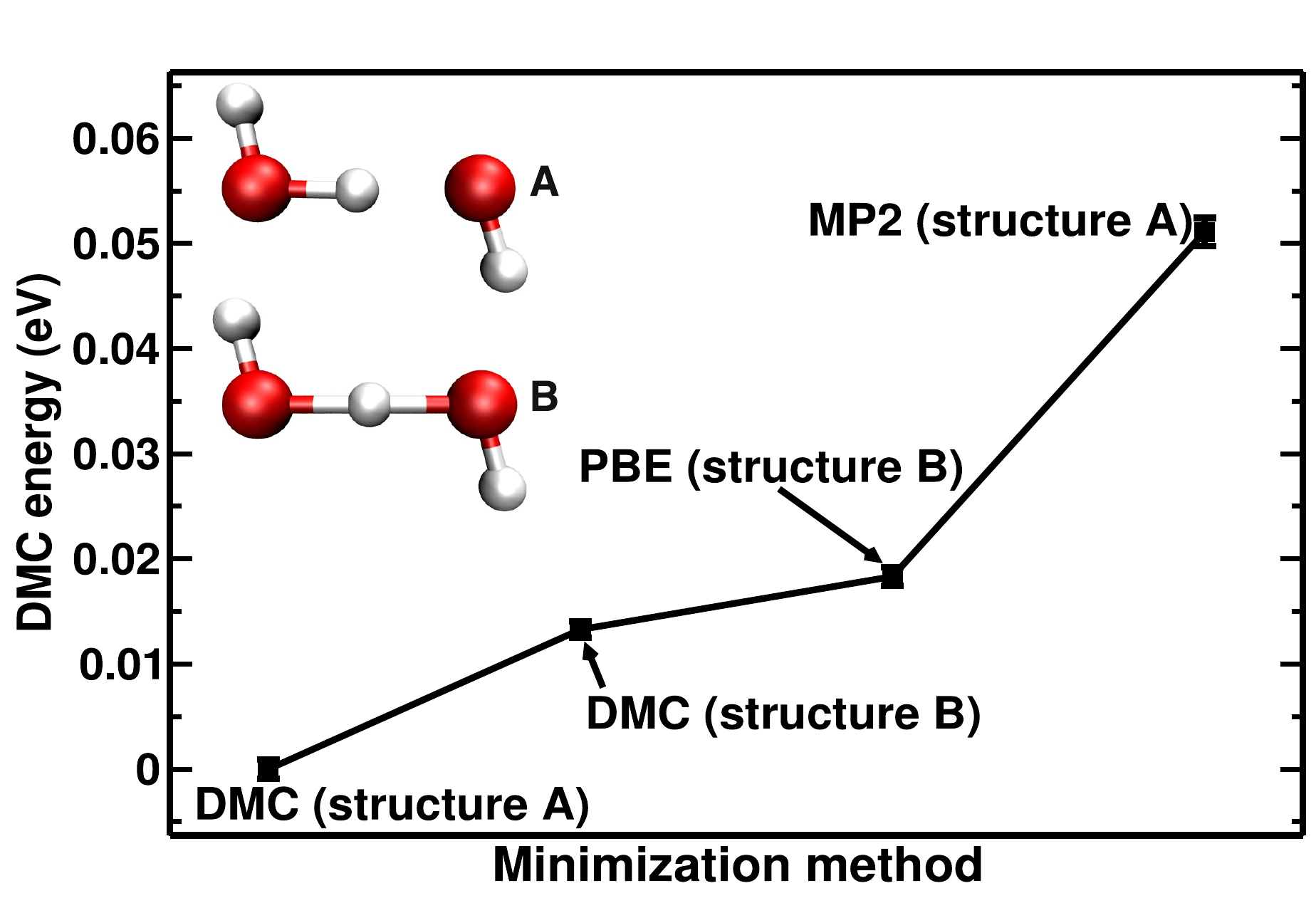}
\caption{(color online) The relative DMC energies of geometries obtained by minimizing different potential energy surfaces.  The line is a guide to the eye.  Stochastic error bars 
are the size of the symbols.  }
\label{fig:pes_scan}
\end{figure}

To show the value of optimizing geometries within QMC,  we apply the method to the H$_2$O-OH$^-$ complex (Fig~\ref{fig:pes_scan}), which is present in liquid water and important in many systems in condensed matter, biology, and chemistry. The shape of this potential energy surface is crucial to understanding hydrogen transfer in water.  In this case, we use our knowledge of the system to choose fixed search directions for efficiency reasons, omitting the Hessian inference method.  For systems that are not as easily decomposed, the Hessian inference is invaluable.
 It has been noted\cite{perdew_h3o2} that current DFT functionals disagree on the ground state structure of this complex.  The potential minima are the
non-centrosymmetric structure (A) and the centrosymmetric structure (B). 

Hartree-Fock and second order M{\o}ller-Plesset
perturbation theory (MP2) find that structure A is lower in energy, with a barrier to transfer the proton.  This is the 
traditional picture of this structure.  The local density approximation and generalized gradient approximation (PBE\cite{pbe}) of DFT find that structure B is lower in energy.   Using our QMC line minimization method without constraints (seven degrees of freedom) we find structure A to be the minimum energy.  The results are summarized in Table~\ref{table:geom}.  DMC differs {\em qualitatively} from the DFT 
results in that structure A is the minimum, and quantitatively with MP2, since the oxygen-oxygen distance
in MP2 is much smaller than in DMC for structure A.

\begin{table}
\caption{H$_2$O'-OH$^-$.  The non-QMC methods are optimized using conjugate gradient routines in GAMESS.
The asterisked DMC result corresponds to constraining the DMC geometry search to the symmetry of structure B. } 
\label{table:geom}
\begin{tabular}{lccc}
%& \multicolumn{2}{c}{\bf Structure A} & {\bf Structure B} \\
{\bf Method} & {\bf O-O' } & {\bf O'-H }  & {\bf Structure type} \\ %&  {\bf O-O'}\\
\hline
LDA & 2.448 & 1.224 & B \\
PBE & 2.470 & 1.235 & B\\
MP2 & 2.469 & 1.123 & A\\
%B3LYP & 2.491 & 1.109 & A \\
%TPSS\cite{perdew_h3o2} & --- &  1.146 & A \\
DMC & 2.491(2) & 1.111(3) & A \\
DMC*  &  2.469(3) & 1.235(2) & B\\
%LDA   & N/A    & N/A    & 2.448 \\
%BLYP  & 2.505    & 1.147    & 2.482\\
%B3LYP & 2.491    & 1.109    & 2.491 \\
%PBE0  & 2.446    & 1.163    & 2.438 \\
%PBE & N/A & N/A &  2.470 \\
%MP2   & 2.469    & 1.123    & 2.408 \\
%DMC   & 2.491(2) & 1.111(3) & 2.469(3)\\
\end{tabular}
\end{table}

In Fig~\ref{fig:pes_scan}, we present the DMC energies of several minima.   Without the geometry optimization algorithm, we would use minima from some other methods, for example, PBE and MP2 to obtain structures B and A respectively, then evaluate the total energy using DMC.  In this case,  since the MP2 approximation obtains a poor O-O distance, DMC predicts a much higher energy for structure A and thus favors the centrosymmetric geometry.  However, this is qualitatively incorrect, as we can see from the DMC-optimized structures, which predict structure B to be about 0.015 eV higher in energy than structure A.  It is notable that 0.015 eV is a very small energy difference on the scale of chemical 
bonding, which is why a small error in the geometry by using a lower level theory is enough to reverse the ordering.
 This energy difference is an upper bound 
to the barrier to transfer the hydrogen, and thus at room temperature the hydrogen is free to transfer between the oxygen atoms.  This may have important implications for the development of effective solvent models for water.

%\section{Conclusion}

 In this work, we have viewed the DMC potential energy surface as a given without attempting to improve the accuracy over a simple Slater-Jastrow form; however, our method can also be used to optimize the nodal surface directly and thus further improve the accuracy of the DMC calculation.  Similarly, any Monte Carlo method that depends on continuous parameters could put the stochastic line minimization algorithm to use.  Stochastic line minimization may find application in experiments whose objective is to obtain a precise minimum in many-dimensional space, but the experiments are difficult to perform precisely.  If it is instead easier to perform many iterations of the experiment, then the minimization method that we have outlined may be applicable.  The scaling for a quadratic potential is ${\cal O}(N_{DOF}^2)$ if only values are used; if even approximate gradients are available, the scaling can be reduced to approximately ${\cal O}(N_{DOF})$, as discussed in the EPAPS document.

%The algorithm that we have outlined is not optimized for finding a minimum when the location is not well-known; however, this is not a significant limitation.  It is only %necessary to employ the averaging near the minimum, where the objective function is flat.  Where the derivative is large, traditional value-based searches such as Powell's line %minimization\cite{powell} or even finite-difference based estimations of the gradient can be employed.

%In conclusion, we have introduced a stochastic line minimization method that uses noisy evaluations of an %objective function to obtain multidimensional minima with error bounds {\bf kind of repetitive}.  As applied %to the problem of using the results of quantum Monte Carlo electronic structure calculations to find minimum %energy geometries, we have demonstrated nine degrees of freedom, with more quite likely.  We have found that %in some cases, the high accuracy QMC minimal energy geometries are qualitatively different from traditional %methods in an important model molecule, H$_2$O-OH$^-$.  

The algorithm is quite general, requiring only evaluations of the objective function with a known distribution, and can thus be applied to many minimization problems.  At its core, it is a method for converting noisy value calculations into precise minima with rigorous error bounds, so it can be applied in any of the many problems where that is necessary.  In electronic structure, diffusion Monte Carlo is accurate and applicable to many systems beyond the ground state of molecules, including solid structures and excited states\cite{schautz_excited}.   The scaling for DMC geometry optimization is then ${\cal O}(N_{DOF}^{1-2}N_e^{2-3})$, depending on the size of the system and whether gradients are available.  This favorable scaling combined with the already known high accuracy of quantum Monte Carlo could open up new levels of accuracy in structure determination in condensed matter and chemical physics.

This work was supported by the Department of Energy under grant No. DE-SC0002623 and the 
 NSF under grant No. 0425914.  We wish to thank NERSC and Teragrid as well for providing computational resources.  We would also like to acknowledge Yosuke Kanai for useful discussions.

\bibliography{geom}

\appendix

\section{Details of the minimization algorithm}

In this EPAPS document, we step through the minimization process  in detail.
Note that some of the details are not necessarily perfectly optimal, but they do work robustly and efficiently.  We thus provide
this description only as a reference implementation that could potentially be improved significantly.

\subsection{Line minimization}

Suppose we wish to minimize a function along a line using the values.  For a deterministic function, a good method is the golden ratio method.  However, for a non-deterministic function, locating the minimum by comparing the value is bound to be inefficient, since the objective function $f(x)$ scales as $f(x)\propto x^2$, and the work to resolve energy differences scales as the inverse square root of the difference.  Therefore, near the minimum, the effort to resolve position differences scales as the inverse {\em quartic} root of the difference.  The computational cost using fitting (and gradients) on the other hand, scales as the inverse square root.  Whether fitting using the value of the function or the gradient of the function is more accurate will depend on their relative variances and the specific problem.  In this work, we use exclusively values since they are easily attainable in quantum Monte Carlo methods.

We developed a reliable strategy to bracket the minimum.  Given a starting position ($t_0$) and a trust radius $r_t$, the algorithm works with three points, initialized as follows:  $t_1=t_0-r_t$, $t_2=t_0$, $t_3=t_0+r_t$.  Let $f_i=f(t_i)+\chi$ be the function evaluated at $t_i$ with a normal random error $\chi$ with standard deviation $\sigma$, which can be controlled by performing longer calculations.  Define $S(f_i,f_j)$ as being true when $f_i$ is significantly different from $f_j$, i.e., about four standard deviations away.  A  basic algorithm as follows:
\begin{enumerate}
\item Evaluate $f_1$, $f_2$, and $f_3$ at uncertainty $\sigma$
\item If $S(f_1,f_2)$ and $S(f_2,f_3)$ : 
\begin{itemize} 
  \item If $f_2< f_3$ and $f_2 < f_1$ stop.  We have bracketed the minimum.
  \item If $f_2 < f_3$ and $f_2 > f_1$, then set $t_i=t_i-r_t$ for $i=1,2,3$, return to 1.
  \item If $f_2 > f_3$ and $f_2 < f_1$, then set $t_i=t_i+r_t$ for $i=1,2,3$, return to 1.
 \end{itemize}
\item Else $\sigma=\sigma/2$ and return to 1.
\end{enumerate}
This algorithm occasionally gets stuck when two points are roughly the same distance from the minimum and thus have nearly the same value.  This can be ameliorated by searching between points when only two are insignificantly different.

Once the minimum is bracketed to within the trust region, we can then fill in points along the line and fit.  The distribution of points is not very critical, so long as there are a few points far away from the minimum to determine the curvature and higher order parameters accurately.  We typically use a four parameter cubic form: $f(x)=c_0+c_2(x-c_1)^2+c_3(x-c_1)^3$, which gives a good balance between being accurate, easy to fit, and few parameters.  A quadratic form requires trust radii that are very small, reducing the efficiency of the algorithm.  We perform the fit using Bayesian inference given the set of data $D=\{t_i,f_i, \sigma_i\}$.  The likelihood function is given by
\begin{equation}
L(\bc|D) \propto p(D|\bc),
\label{eqn:line_likelihood}
\end{equation}
% $p(\bc)$ is the prior distribution, which we set to 1, and $p(D)$ is a normalization constant.  We evaluate the probability distribution function as 
where
 \begin{equation}
 p(D|\bc) \propto \prod_i \exp[-(f(t_i, \bc)-f_i)^2/2\sigma_i^2]
 \end{equation}
is the probability that we would have obtained the data given a set of parameters.  The distribution of a given parameter $c_i$ can be found by taking the marginal distribution of $L(\bc|D)$, or for more speed at the cost of not knowing the uncertainty well, simply using the maximum likelihood estimator.

\subsection{Hessian estimation}

 One begins by noting that near the minimum, the energy is described by a quadratic function:
\begin{equation}
E(\bx) = \delta \bx^T H \delta \bx + E_0.
\end{equation}
Far away from the minimum, this is an approximation to speed up the minimum search.  Near the minimum, we used this property to prove the average properties of the stochastic sequence of line minimizations.  Recall that only the minimum of the line minimization must be in the quadratic region.  With no further assumptions, we can estimate the Hessian (and full quadratic surface) given a collection of line minimizations, even in the presence of stochastic noise.

Let $\bc_Q$ be the collection of all parameters for the quadratic region, $\bc_\ell$ the collection of line parameters for line $\ell$. We wish to find $p(\bc_Q | D)$, where D is the set of function evaluations, arranged in lines.  By Bayes' theorem, we know that
\begin{equation}
 p(\bc_Q |D ) \propto p(D|\bc_Q) = \prod_\ell \int p(D_\ell | L_\ell) p(L_\ell | \bc_Q) d \bc_\ell 
\end{equation}
Since $p(L_\ell | \bc_Q)$ is not based on stochastic data, it is a delta function that forces $\bc_\ell$ to be consistent with $\bc_Q$ as follows.  In the quadratic region, $E(\bx)=\delta \bx^T H \delta \bx +E_0$, minimizing along a direction $\bv$ from a starting position $\bx_0$ gives the one-dimensional function 
\begin{equation}
E(t)=(\bx_0+t\bv-\bm)^T H (\bx_0+t\bv-\bm)+E_0.
\end{equation}
This gives the following constraints on each set of line parameters $c_\ell$: 
\begin{equation}
t_m=- \frac{ \bv^T H (\bx_0-\bm) + (\bx_0-\bm)^T H \bv }{ 2 \bv^TH\bv }
\end{equation}
\begin{equation}
\left.\frac{d^2E}{dt^2} \right|_{t_m} = 2 \bv^T H \bv
\label{eqn:curve_constraint}
\end{equation}
\begin{equation}
E(t_m)= (\bx_0+t_m\bv -\bm)^T H (\bx_0 + t_m \bv - \bm)
\end{equation}
We thus only need to integrate over the variables in each line that are not specified by the above equations.  We can replace the $\bc_\ell$ by   the reduced line parameters $\tilde \bc_\ell$.  Since there are three constraints per line, the dimensionality of $\tilde \bc_\ell$ is three less than the dimensionality of $\bc_\ell$.  We can recover the full set of line parameters as a function of the quadratic parameters and the reduced line parameters; i.e., $\bc_\ell (\bc_Q, \tilde \bc_\ell )$

We eventually want the likelihood function for the quadratic parameters and the reduced line parameters: $L(\bc_Q, \{\tilde \bc_\ell \} | D)$, where $D$ is the data.  This is given by 
\begin{equation}
L(\bc_Q, \{\tilde \bc_\ell \} | \{D_\ell\} ) \propto \prod_\ell L(\bc_\ell(\bc_Q, \tilde \bc_\ell) | D_\ell ) .
\label{eqn:quadratic_likelihood}
\end{equation}
where $L(\bc_\ell|D_\ell)$ is given by Eqn~\ref{eqn:line_likelihood}.

The first estimate of the Hessian matrix is achieved by minimizing the violation of Eqn~\ref{eqn:curve_constraint} from the line fits.  The minimum of the quadratic model is estimated as the last minimum found in a line minimization, and the minimum energy similarly.  The parameters are then optimized for maximum likelihood using a conjugate gradient optimization method.  Finally, the parameters are randomized and reoptimized to get out of local maxima. We find that the maximum likelihood estimator is typically quite sufficient for determining search directions; it is not necessary to sample to find distributions.

Note that each line minimization places three constraints on $\bc_Q$ and $\bc_Q$ has a dimensionality of $1+N_{DOF}+N_{DOF}(N_{DOF}+1)/2$, so to make this a fully determined problem, $1/6+ N_{DOF}/3 + N_{DOF}(N_{DOF} +1)/6$ line minimizations are needed.  This is always smaller than the $N_{DOF}^2$ line minimizations necessary in Powell's method, the standard in gradient-free minimization, with the additional benefit that the Hessian inference method is uncertainty-aware, which allows much higher efficiency when uncertainty can be traded for computational time. 

If gradient information is available, then they can be integrated directly into the likelihood function from Eqn~\ref{eqn:quadratic_likelihood} as a simple product:
\begin{align}
L(\bc_Q, \{\tilde \bc_\ell \} | \{D_\ell\} , \{\bg_i\}) \propto  & \\ \notag
\left[ \prod_\ell L(\bc_\ell(\bc_Q, \tilde \bc_\ell) | D_\ell ) \right] & \left[ \prod_i L(\bc_Q | \bg_i) \right]
\end{align}
where the likelihood function can be approximated as 
\begin{align}
L(\bc_Q | \bg_i) \propto &  \\ \notag
\exp [ - (\nabla_{\bx} & E(\bx_i, \bc_Q) - \bg_i)^T C_i  (\nabla_{\bx} E(\bx_i, \bc_Q) - \bg_i) ],
\end{align}
and $C_i$ is the covariance matrix between all the gradient components.  The Hessian inference then becomes essentially a quasi-Newton algorithm that is uncertainty aware.  Approximate gradients are also useful, since minima can be determined by using only the value to perform the line fits.  In some instances, one can imagine elaborations of this where DFT or other approximations to the gradient are used to accelerate the minimum search.

\subsection{A detailed look as the algorithm proceeds}

\begin{figure}
\includegraphics[width=\columnwidth]{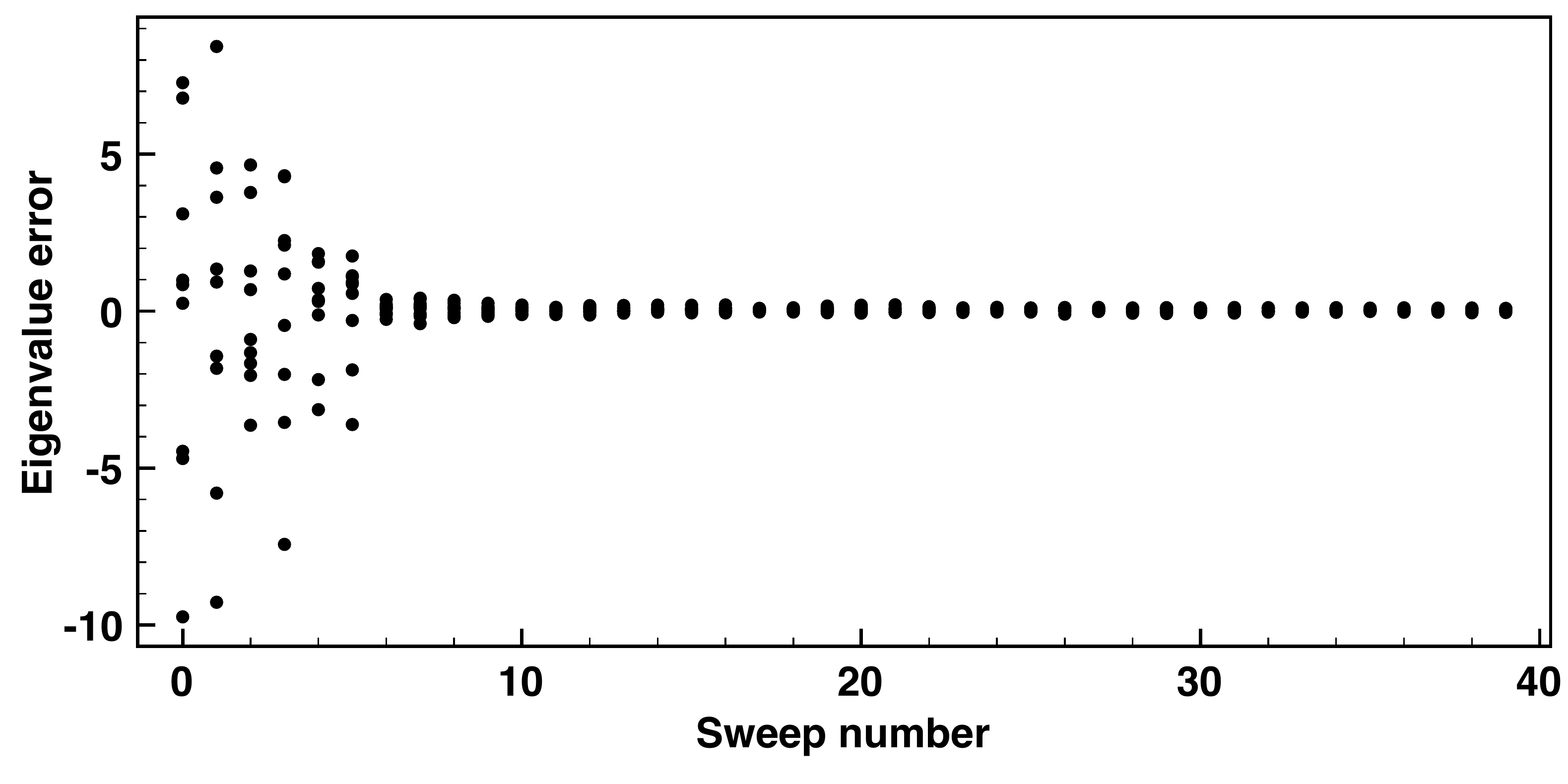}
\includegraphics[width=\columnwidth]{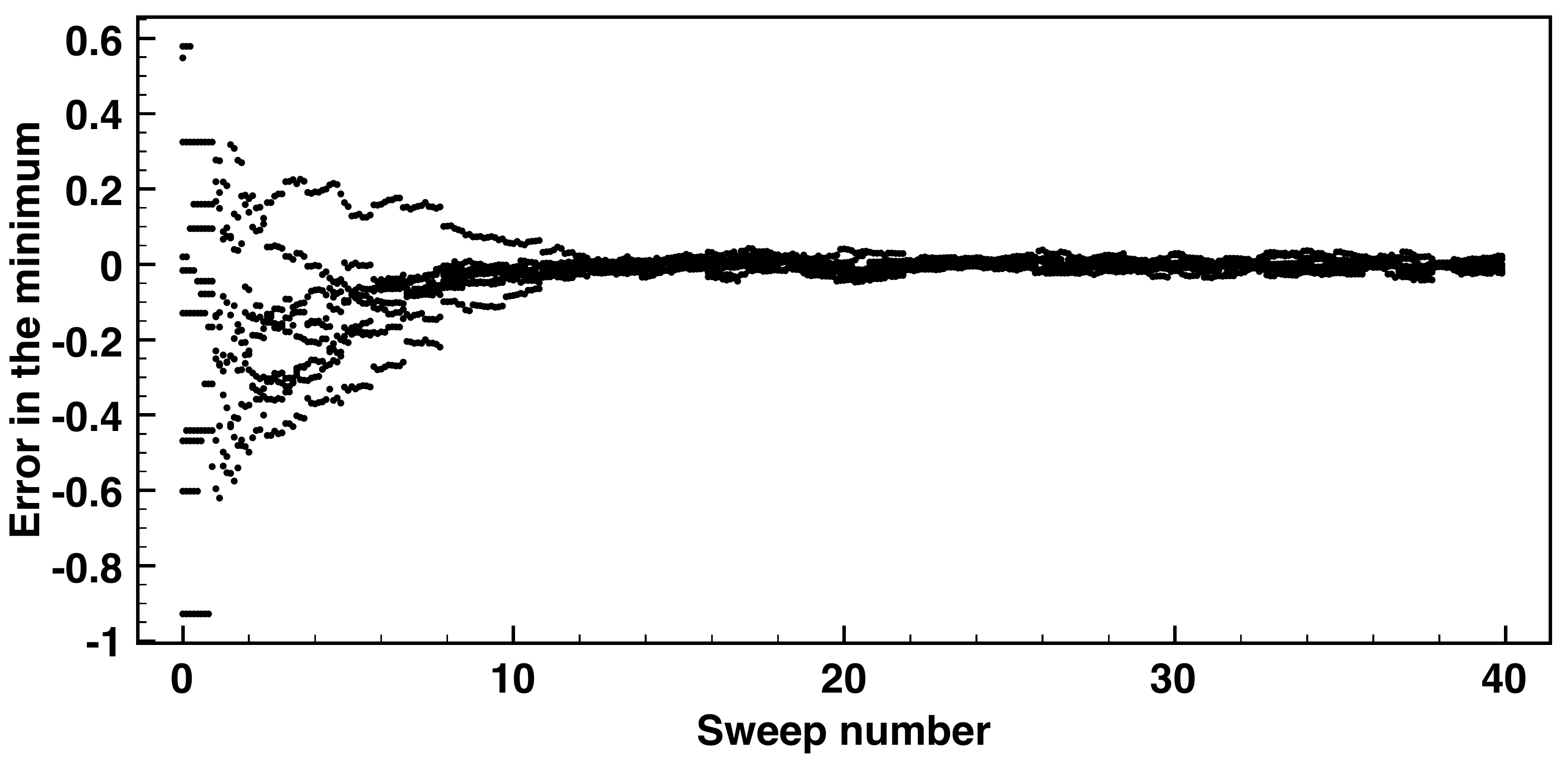}
\caption{Typical convergence of the eigenvalues and the estimate of the minimum for a nine-dimensional optimization problem. The potential energy surface was a random quadratic function with a cubic perturbation.  The potential energy surface was calculated as $E(\bx)+\chi$, where $\chi$ was a random Gaussian variable with mean zero. }
\label{fig:converge}
\end{figure}

\begin{enumerate}
\item Start a line minimization sweep, for each direction
  \begin{itemize}
    \item Bracket the minimum
    \item Fill in data points within the trust region from the minimum
     \item Fit the data points with a function
    \item Update the minimum
    \item Add fit to list of fits
    \end{itemize}
\item Prune list of fits to contain only the last $N_{history}$ sweeps
\item Fit N-dimensional quadratic model to lines, find maximum likelihood
\item Diagonalize Hessian
\item Update directions
\item Return to 1.
\end{enumerate}

In Fig~\ref{fig:converge}, we show the convergence of the eigenvalues from the Hessian estimation to the exact ones, and the minima from the line minimizations. 

\section{Proofs of the properties of the stochastic sequence}

We have the following equations:
\begin{align}
\delta x_i^{(n)}=& \chi_i^{(n)}-\sum_{j \neq i} \frac{H_{ij}}{H_{ii}} \chi_j^{(n-1)} \notag \\
+&\sum_{j \neq i} \sum_{k \neq j} \frac{H_{ij}H_{jk}}{H_{ii}H_{jj}} \chi_k^{(n-2)}+\ldots
\label{eqn:avg_app}
\end{align}
\begin{align}
\operatorname{Var}(\delta x_i^{(n)}) & =  \operatorname{Var}(\chi_i^{(n)})  
+\sum_{j \neq i} \frac{H_{ij}^2}{H_{ii}^2} \operatorname{Var}(\chi_j^{(n-1)}) \notag \\
+& \sum_k  \operatorname{Var}(\chi_k^{(n-2)})\left[\sum_j (1-\delta_{ij})(1-\delta_{jk})\frac{H_{ij}H_{jk}}{H_{ii}H_{jj}}\right]^2 \notag \\
+& \ldots
%\sum_{j \neq i} \sum_{k \neq j} \frac{H_{ij}^2H_{jk}^2}{H_{ii}^2H_{jj}^2} \operatorname{Var}(\chi_k^{(n-2)})+\ldots
\label{eqn:var}
\end{align}
Suppose that $\sum_{j \neq i} \left| H_{ij}/H_{ii} \right| \le x, \forall i$ for some $0 < x < 1$.   
Then $|\sum_{k \neq j} \frac{H_{ij}H_{jk}}{H_{ii}H_{jj}} | \le \left|\frac{H_{ij}}{H_{ii}}\right|x$, $|\sum_{j \neq i} \sum_{k \neq j} \frac{H_{ij}H_{jk}}{H_{ii}H_{jj}} | \le x^2$, and so on.

\subsection{The average is equal to the true minimum}
Suppose that $\langle \chi_i^{(j)} \rangle=0 \forall i,j$.  Then 
\begin{equation}
\langle \delta x_i^{(n)} \rangle \le x^n \delta x_j^{(0)}.
\end{equation}
This goes to zero exponentially in $n$, so we can obtain an average arbitrarily close to zero by starting the average at high enough $n$.

\subsection{The average has finite variance} 

Assume $\operatorname{Var}(\chi_i) \le v_m \forall i$.  Then
\begin{align}
\operatorname{Var}(\delta x_i^{(n)})  %\le&  v_m+x^2v_m +x^4v_m + \ldots  \notag \\
\le v_m(1+x^2+x^4+\ldots) =  \frac{v_m}{1-x^2},
\label{eqn:var2}
\end{align}
for large $n$.

\subsection{The sequence has a finite correlation time}
Suppose that 
\begin{equation}
\langle \chi_i^{(k)} \chi_i^{(j)} \rangle = \sigma^2 \delta_{jk}  \forall i,j,k
\end{equation}
for some variance $\sigma^2$.  
Then the correlation between elements $n$ and $j$ in the sequence is
\begin{equation}
|\langle \delta x_i^{(j)} \delta x_i^{(n)} \rangle | \le \sigma^2 \sum_{k=0}^j x^{n-k} =  \sigma^2 x^{n-j} \sum_{k=0}^j x^k,
\end{equation}
where we reversed the sum indices.  As $n$ and $j$ become large while keeping the difference finite, we can then find that 
\begin{equation}
|\langle \delta x_i^{(j)} \delta x_i^{(n)} \rangle | \le \frac{\sigma^2 x^{n-j}}{1-x}.
\end{equation}
This goes to zero exponentially with $n-j$, and since the variance is finite, the correlation coefficient also goes to zero exponentially.

\end{document}